\documentclass[10pt, conference, compsocconf]{IEEEtran}

\pdfoutput=1

\usepackage{color}
\usepackage{xspace}
\usepackage{amsmath}
\usepackage{alltt}
\usepackage{listings}
\usepackage{tikz}
\usepackage{pgfplots}
\usepackage{pgfplotstable}
\usetikzlibrary{pgfplots.groupplots}
\usepackage{pdfpages}
\usepackage{xcolor,colortbl}

\definecolor{lightred}{rgb}{1,0.85,0.85}
\definecolor{lightblue}{rgb}{0.85,0.85,1}

\newcommand{\fig}[1]{Fig.~\ref{#1}\xspace} 
\newcommand{\tbl}[1]{Table~\ref{#1}\xspace} 
\newcommand{\secref}[1]{Section~\ref{#1}\xspace} 
\newcommand{\eq}[1]{(\ref{#1})\xspace} 

\newcommand{\daint}{Piz Daint\xspace} 
\newcommand{\todi}{T{\"o}di\xspace} 
\newcommand{\rosa}{Monte Rosa\xspace} 
\newcommand{\clogin}{clogin\xspace} 

\newcommand{\watt}[1]{{#1}~W\xspace} 
\newcommand{\joule}[1]{{#1}~J\xspace} 
\newcommand{\hertz}[1]{{#1}~Hz\xspace} 

\newcommand{\pmcounters}{pm\_counters\xspace} 
\newcommand{\nops}{n_{\tiny\mbox{ops}}} 

\begin{document}
%
\title{First Experiences With Validating and Using the Cray Power Management Database Tool}


\author{
  \IEEEauthorblockN{
    {Gilles Fourestey\IEEEauthorrefmark{1}},
    {Ben Cumming\IEEEauthorrefmark{1}},
    {Ladina Gilly\IEEEauthorrefmark{1}},
    and
    {Thomas C. Schulthess\IEEEauthorrefmark{1}\IEEEauthorrefmark{2}\IEEEauthorrefmark{3}}\\
  }
  \IEEEauthorblockA{\IEEEauthorrefmark{1} Swiss National Supercomputing Center, ETH Zurich, 6900 Lugano, Switzerland}
  \IEEEauthorblockA{\IEEEauthorrefmark{2} Institute for Theoretical Physics, ETH Zurich, 8093 Zurich, Switzerland}
  \IEEEauthorblockA{\IEEEauthorrefmark{3} Computer Science and Mathematics Division, Oak Ridge National Laboratory, Oak Ridge, TN 37830, USA}
    Email: \{gilles.fourestey, ben.cumming, ladina.gilly, thomas.schulthess\}@cscs.ch
}

\maketitle

\begin{abstract}
In October 2013 CSCS installed the first hybrid Cray XC-30 system, dubbed Piz Daint. This system features the power management database (PMDB), that was recently introduced by Cray to collect detailed power consumption information in a non-intrusive manner. Power measurements are taken on each node, with additional measurements for the Aries network and blowers, and recorded in a database. This enables fine-grained reporting of power consumption that is not possible with external power meters, and is useful to both application developers and facility operators. This paper will show how benchmarks of representative applications at CSCS were used to validate the PMDB on Piz Daint. Furthermore we will elaborate, with the well-known HPL benchmark serving as prototypical application, on how the PMDB streamlines the tuning for optimal power efficiency in production, which lead to Piz Daint being recognised as the most energy efficient petascale supercomputer presently in operation.

\end{abstract}

\section{Introduction}
\label{sec:introduction}
%
Traditionally the sole performance concern of high-performance computing was time to solution of a computation. This metric lead to the dominance of the high-performance Linpack (HPL) benchmark which, in the 90's, formed the basis for the top500 \cite{netlib:website}, \cite{LINPACKuserguide}, which ranks application performance based on high arithmetic intensity on a specific system. HPL is a benchmark that performs a LU factorization and a triangular matrix solve of a dense linear system and relies heavily on the high arithmetic intensity of the BLAS-3 matrix matrix multiplication kernel DGEMM. 

Looking at the top500 list and its trend since its introduction in 1993, HPL performance has increased by a factor of roughly 1000 every 10 years: in November 1993, the Connection Machine CM-5/1024 was ranked number 1 with a Rpeak of 131 GFlops, and as of November 2013 the fastest supercomputer was Tianhe-2 with Rpeak of 54.9024 PFlops: a factor of 420'000 over 20 years.

As the peak performance of supercomputers has increased over time, another metric has become more important: power consumption. In June 2008 the first petaflop system was introduced with a power consumption of 2.3 MWatts. Five years later in June 2013, the fastest supercomputer delivered 33 Tflops using 17.8 MWatts. If this trend continues, the first exascale system would require above 100 MWatts.

In 2005, the Green500 \cite{green500:website} was created in order to shift the focus from speed, and therefore Top500, to the Gflops/W metric to promote energy efficiency and reliability. However, controversy has appeared concerning the energy measurements techniques used as they where somehow ill-defined and prevented reproducibility, leading to detailed procedure published by the Energy Efficiency HPC working group this year (see \cite{eehpcwg:website} and \cite{scogland2014} for more details).

To quantify energy efficiency, one must measure power consumption. This can be a difficult and tedious process, especially on large HPC systems. However, to effectively optimize an application for energy efficiency, power consumption should be easily available to developers, like performance counters or time to solution.

In the XC-30 line, Cray introduced a new tool that makes power and energy information readily available.
This tool is the Power Management Data Base (PMDB), a PostgreSQL database available with Cray SMW, that collates power measurements from a comprehensive set of hardware sensors throughout the system.
With the PMDB, it is possible to  access the power consumption, energy, current and voltage of the racks, blades and nodes of an XC-30 system through simple database queries or through polling specific system files.

The goal of the present contribution is to validate this capability at the application level, both in terms of usability and accuracy.

\section{Overview}
\label{sec:overview}
We will give an overview of the level-3 power measurement capabilities that are integrated into the facilities at the Swiss National Supercomputing Center (CSCS) in \secref{sec:facility}.
The power monitoring facilities that are integrated into the XC-30 platform will outlined in \secref{sec:methodology}, with a discussion of how these can be used to determine total energy consumption of an application.
\secref{sec:hpl} will use a hybrid implementation of the HPL benchmark to show how we tuned and optimized both performance and energy consumption easily using PMDB, with a summary of the tuning process used for the Green500 results.
A selection of three real world applications will be used in \secref{sec:cosmo} and \secref{dca} to validate power measurements using the external power meter and the PMDB.
We will conclude in \secref{sec:conclusion}.

\section{Setup for Level 3 Facility-Side Power Measurements}
\label{sec:facility}
\begin{figure}[htp!]
\centering
\includegraphics[width=0.32\textwidth]{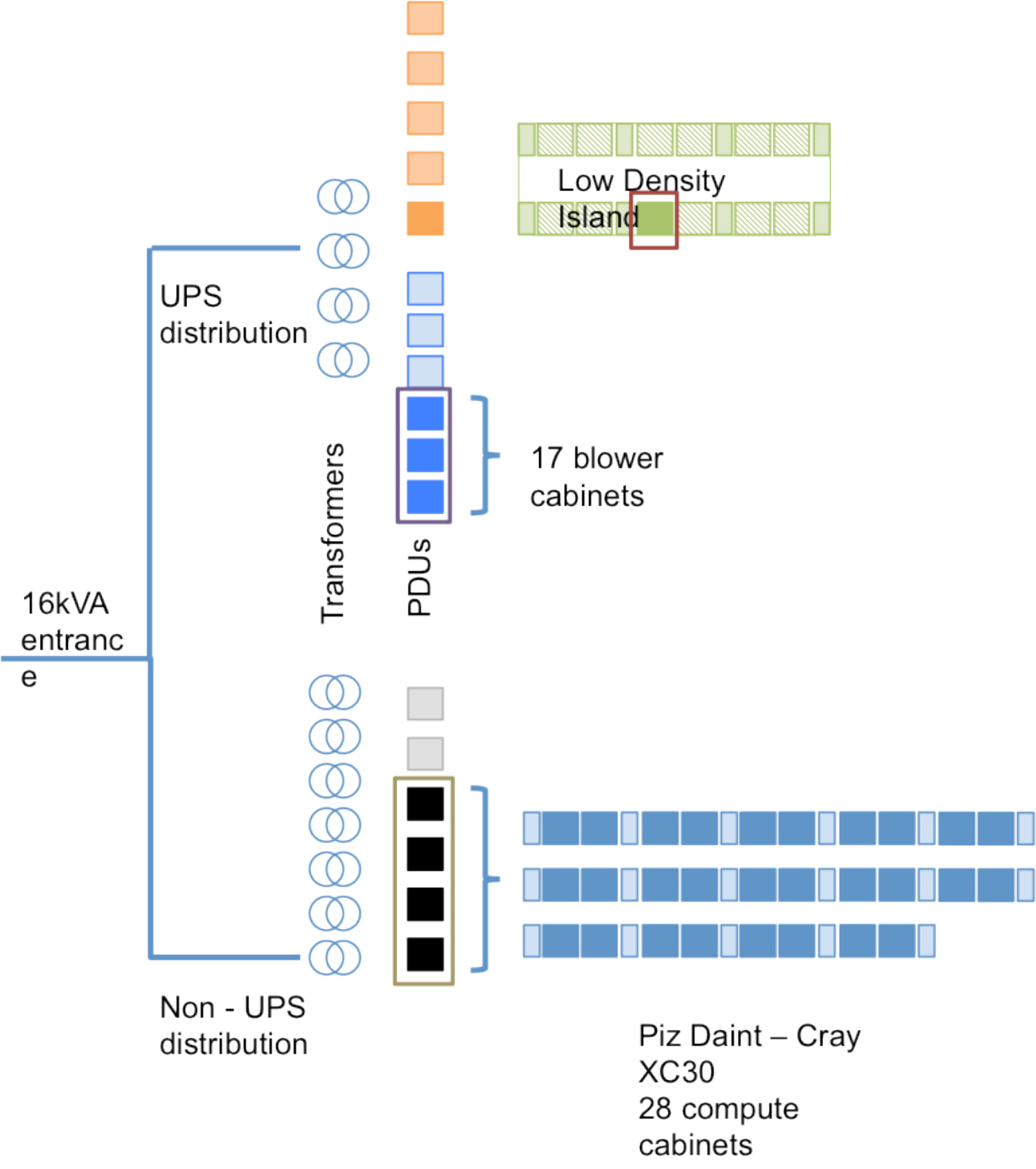}
\includegraphics[width=0.15\textwidth]{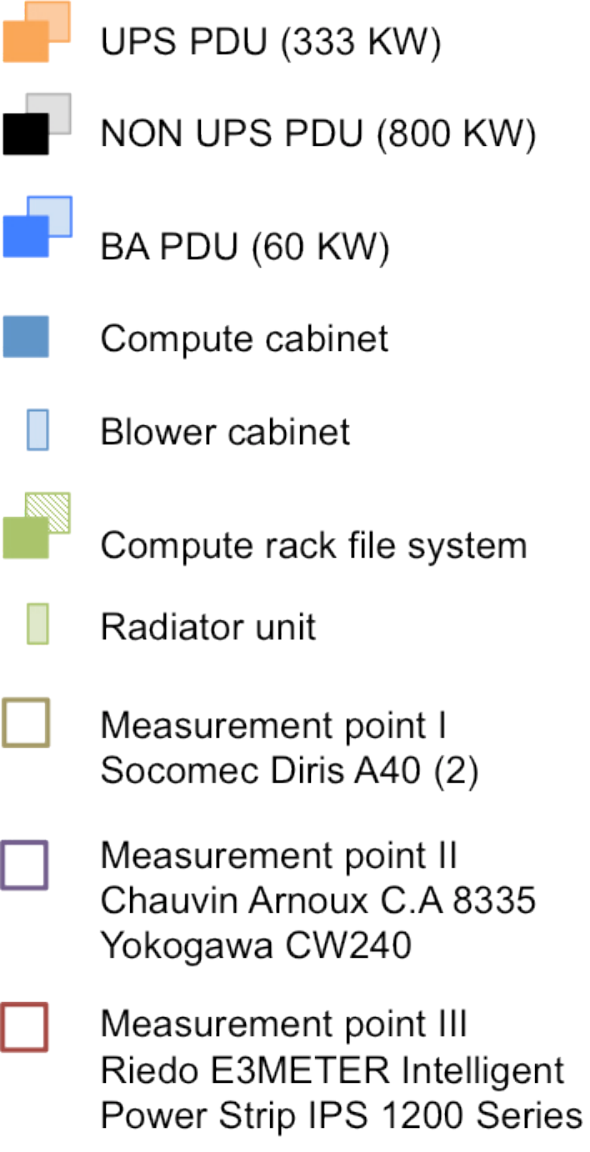}
\caption{Facility level power measurement location and instruments for Piz Daint system at CSCS.}
\label{fig:facilitymeasurements}
\end{figure}

CSCS infrastructure provides both utility power and UPS power to the machine room, with UPS power only connected to critical infrastructure. \fig{fig:facilitymeasurements} shows the power measurement location and instruments for our XC-30 system Piz Daint.

In the case of the Piz Daint Cray XC-30 machine power supply is different for the following three component sets:
\begin{itemize}
    \item   Compute cabinets (utility power)
    \item   Blower cabinets (UPS power)
    \item   File system (utility power)
\end{itemize}

The compute cabinets are connected to utility power and are distributed over 4 power distribution units (PDUs). The blower cabinets are connected to UPS supply from 3 PDUs that also provide power to the pumps for the cooling circuits. The file system is housed within one rack of an in-row cooling island where it is connected to UPS power supply via in-rack PDUs.

The compute cabinets are connected to 4 PDUs. Measurements were taken directly from the electrical network management system that does a continuous energy reading within the PDUs. These measurements are taken with a Socomec Diris A 40(2) device classified as Class 1 precision. The instruments were calibrated in November 2011.

The blower cabinets are connected to 3 different PDUs on the UPS power supply. These PDUs also supply the power to the pumps for the Piz Daint cooling loop. In order to be able to take a precise reading for the blower cabinets the pumps where set to a fixed regime, thereby also fixing their power consumption to 16kW so this could be deducted from the measurements to obtain the power consumption for the blower cabinets. These PDUs are not equipped to provide continuous energy readings to the network management system, so the measurements for the blower cabinets had to be taken with power meters attached to the PDU power supply. The devices used where a Chauvin Arnoux C.A 8335, clamp model MN93A with a precision of +/- 0.7\% as well as a Yokogawa CW2 with a precision of +/- 0.5\%. Both instruments had been calibrated on November 9th 2012. These instruments measure instantaneous power at a frequency of once per second.

The file system is located within one rack of an in-row cooling island. It is connected to UPS power supply via in-rack Riedo PDUs. The measurements for the file system were extracted from the Riedo E3METER Intelligent Power Strip IPS 120 Series that have a precision of 1\%. These instruments were calibrated in March 2011.

\section{Software Power Measurements}
\label{sec:methodology}
\subsection{Energy/power consumption software measurements}
The \daint XC-30 platform features advanced hardware and software features for monitoring energy consumption, that utilize hardware sensors monitor throughout the system.
Each XC-30 blade features hardware sensors, which are used to provide total power consumption and GPU power consumption on each node.
Access to this power consumption data is available to the user via three different mechanisms, which we will cover in more detail below.
\subsection{PMDB}
Information from the sensors is collected by the Power Measurement Data Base (PMDB), which is discussed in more detail in~$\cite{craypmdb}$.
The PMDB is a database with comprehensive power readings at a \hertz{1} frequency for every node, GPU, blower and network chip.
The database can be queried to provide time series data for any set of nodes, network chips and blowers used by a job.
There are two main ``resolutions'' at which energy information is provided by the database: by cabinet and by node.
The per-cabinet output includes the power consumption of the interconnect, which is the sum of power consumption from the aires network chips on each node in a cabinet, which are measured separately to the node-level measurements.
The power and energy consumption of a job can then easily be derived from a node-level query given the ALPS jobid or apid with simple scripts which will determine the nodes involved and the time stamps of the runs.

The main drawback with the PMDB is that it is hosted on the software management nodes, which are not available to normal users. To get information from the PMDB, users must present a system administrator with a list of nodes and a time stamp, or a job id, which make this approach inconvenient for tight integration into the development process.
\subsection{Ressource Utilization Report}
The two remaining methods are available in user space. The resource utilization report (RUR) for a job provides the total energy consumed by each application run on a compute node. The RUR reports are stored in files that can be searched given the apid (unique id given to each mpi task run on the system), for example to find the energy consumed for the job with apid 2371227:
\\\texttt{\footnotesize
>grep 2371227 /scratch/daint/RUR/* | grep energy
\\...cmdname: ./cosmo energy ['energy\_used', 159028]}\\
which gives \joule{159028} for the call to cosmo. This method is extremely useful in order to get quick power and energy consumption informations for a given job since this information will be available right after the end of each run. However, it is not possible to measure a specific part of the code. In order to do so, a third option is available.\\
\subsection{PM counters}
The \pmcounters are another feature that makes detailed power and energy information available to users while their application rus.
A snap shot of the current power and cumulative energy consumption on each node are provided by \pmcounters Sysfs files\footnote{Sysfs is a virtual file system provided by Linux that exports information about devices and drivers from the kernel device model to user space.}. According to the PMDB manual, the files are:

\begin{center}
\begin{tabular}{r p{6cm}}
    \textbf{power :}      & Point-in-time power in watts \\
    \textbf{energy :}     & Accumulated energy in joules \\
    \textbf{generation :} & A counter increments each time a power cap value is changed \\
    \textbf{startup :}    & Startup counter\\
    \textbf{freshness :}  & Free running counter that increments at a rate of approximately 10 Hz \\
    \textbf{version :}    & Version number for power management counter support \\
    \textbf{power\_cap :} & Current power cap limit in watts; 0 indicates no-capping \\
\end{tabular}
\end{center}

These contents of the files are very easy to understand, for example
\\\texttt{\footnotesize
> cat /sys/cray/pm\_counters/energy\\
18328219 J\\
> cat /sys/cray/pm\_counters/power\\
43 W
}\\
shows that for the node on which the file was polled, the cumulative energy used by the node is \joule{18328219}, and power consumption is \watt{43} (including GPU).

The \pmcounters register files are updated at \hertz{10}, however polling them causes a system interrupt that may affect user code, so their use is not recommended for high-frequency performance monitoring.
For example, they are polled from within the COSMO application to determine the cumulative energy consumption before and after the main time stepping loop, which takes in the order of minutes to run.

Additional sensors monitor the power consumption of the Aires network chip and the blowers. The Aires network has an almost constant power consumption of 100W per rack (or \watt{25} per node) regardless of network load, and the blower power consumption varies from \watt{4400}--\watt{5600}, according to the cooling requirements, though it only exceeds \watt{4400} under very heavy loads on \daint.
\subsection{AC/DC conversion rate}
The power and energy data reported by the sensors do not account for losses during the conversion from alternating current (AC) to direct current (DC).
The efficiency of the conversion from AC to DC is a nonlinear function of power, however an efficiency of 95\% has been verified with an external power meter for a range of test applications.
For example, tests performed with Cray to compare the internal power measurements with those from an external power meter measuring DC current at the wall for an ensemble run of COSMO-2 on one cabinet of hybrid XC-30 agreed almost exactly with an assumed efficiency of 95\% (see \tbl{tbl:cosmoValidate}).
To estimate the total power consumption of a job, given the nodal energy consumption $E_n$ for the $N$ nodes, the total energy consumption $E_j$ for a job can be estimated using the following formula
\begin{equation}
    \label{eq:energyEstimate}
    \text{total energy} = \frac{E_n + N/4\times100\times\tau}{0.95},
\end{equation}
where $\tau$ is the wall time for the application. Equation~\eq{eq:energyEstimate} does not include the contribution of the blowers, whose fan speed (and thus energy consumption) varies between 80\% to 100\%, according to the temperature of the system. In our experiments, we have only observed the blowers going over 80\% for the top500 HPL runs, which used both CPU and GPU at very close to peak. To include the contribution of blowers in results for ``normal'' workloads, the following formula can be used
\begin{equation}
    \label{eq:energyEstimateWithBlower}
    \text{total energy} = \frac{E_n + N/4\times100\times\tau + B \times 4440 }{0.95},
\end{equation}
where $B$ is the number of blowers. The value of $B$ has to be estimated by assigning a portion of the global blower consumption for the system to the nodes in a job. On \daint the value of $B$ for one cabinet is assumed to be 17/28, because there are 28 cabinets and 17 blowers in the entire system. For high-power consuming jobs one might consider using 100\% as an upper limit, or check the PMDB directly for blower energy consumption if available to see if the blower every go above 80\% fan speed.

%
%
%

\section{HPL}
\label{sec:hpl}
\subsection{Problem Setup}


The High Performance Linpack benchmark (HPL) performs an dense matrix LU factorization that relies heavily on the level-3 BLAS kernel DGEMM.
DGEMM has arithmetic intensity is $O(N)$, which means that the number of floating point operations per memory load scaled by the problem size $N$.
The number of floating point operations ($\nops$) performed is $\frac{2}{3}N^3$ and the arithmetic intensity is
\begin{equation*}
\alpha = \frac{\nops}{\mbox{number of load/stores}} = O(N).
\end{equation*}
Hence, for large problems the amount of work done is can be quantified by the total number of floating point operations. 

If we consider that a system has a $K_{\tiny{\mbox{flop}}}$ rate capacity (in FLOPS), then the time to solution (TTS) can be approximated with the following formula
\begin{equation}\label{TTS}
\mbox{TTS} \approx \frac{\nops}{K_{\tiny{\mbox{flop}}}}
\end{equation}
Therefore, minimizing time to solution for a given problem size is equivalent to maximizing the number of FLOPS. 

We define energy to solution (ETS) as the total amount of electrical energy consumed to reach the solution. Given the average power used during the run, $\bar{W}$, then ETS can be expressed as
\begin{equation}\label{ETS}
\mbox{ETS}=\int_0^{T} \bar{W} dt = \bar{W}\times\mbox{TTS}.
\end{equation}
Using ($\ref{TTS}$), we get the following approximation for ETS
\begin{equation}
    {\label{ETS}}
    \mbox{ETS}=\bar{W}\times\frac{\nops}{K_{\tiny{\mbox{flop}}}}.
\end{equation}

Minimizing ($\ref{TTS}$) for a given problem size or $\nops$ is equivalent to maximizing $K_{\tiny{\mbox{flop}}}$, and
minimizing energy to solution is equivalent to maximizing the value of \eq{ETS}, which has units of FLOP/Joule or GFLOPS/W.
Thus it is a common practice to considering that systems with highest FLOPS and FLOPS/W in HPL rely on a metric that is intimately linked to the dense linear algebra motif. Or to put it differently, applications that rely on dense linear algebra like HPL, maximising GFLOPS/W at any given $N$ is equivalent to minimising TTS directly. 

Modern CPUs and GPUs offer very different energy efficiency performances: the theoretical peak performance of the E5-2670 Sandy Bridge CPU found in \daint is 166.4 GFLOPS for a power consumption of 115W, about 1.44 GFLOPS/W, whereas the theoretical peak performance of the K20X GPU is 1311 GFLOPS for a power consumption of 225W, giving 5.83 GFLOPS/W. Therefore, in order to minimize ETS, the CPU should be involved as little as possible, as opposed to the classic HPL benchmark which uses the TTS metric.
\subsection{Intel's P states}
With the introduction of the Sandy Bridge processors, intel provided a scaling driver with an internal governor. 
All the logic for selecting the current P state is contained within the driver and no external governor is used by the cpufreq core. New sysfs files for controlling the P state selection have been added to
$$
\mbox{/sys/devices/system/cpu/}
$$
and it is possible to check the list of possible P states for each core:
\texttt{
\newline
\footnotesize
\textbf{> cat /sys/devices/system/cpu/cpu0/cpufreq/}\\
scaling\_available\_frequencies\\
2601000 2600000 2500000 2400000 2300000 2200000 2100000 2000000 1900000 1800000 1700000 1600000 1500000 1400000 1300000 1200000
} 

Note, that setting a specific P state will prevent the CPU from being overclocked with the Turbo Boost feature, except for the 2601000 frequency.
Cray has implemented the possibility to modify the P state in the user space with aprun command. From the aprun man page:

\textit{Specify the p-state (i.e. CPU frequency) used by the compute
node kernel while running the application. The list of
available frequencies can be found by running the following
command on a compute node:}
\\
\texttt{\footnotesize cat /sys/devices/system/cpu/cpu0/cpufreq\\/scaling\_available\_frequencies}
\\
\textit{If the specified p-state does not match one of the available frequencies, the specified p-state will be rounded down to the next lower supported frequency. Note that the -{}-p-governor and -{}-p-state options cannot be used together, as specifying a p-state implies the use of the ``userspace"
performance governor.}
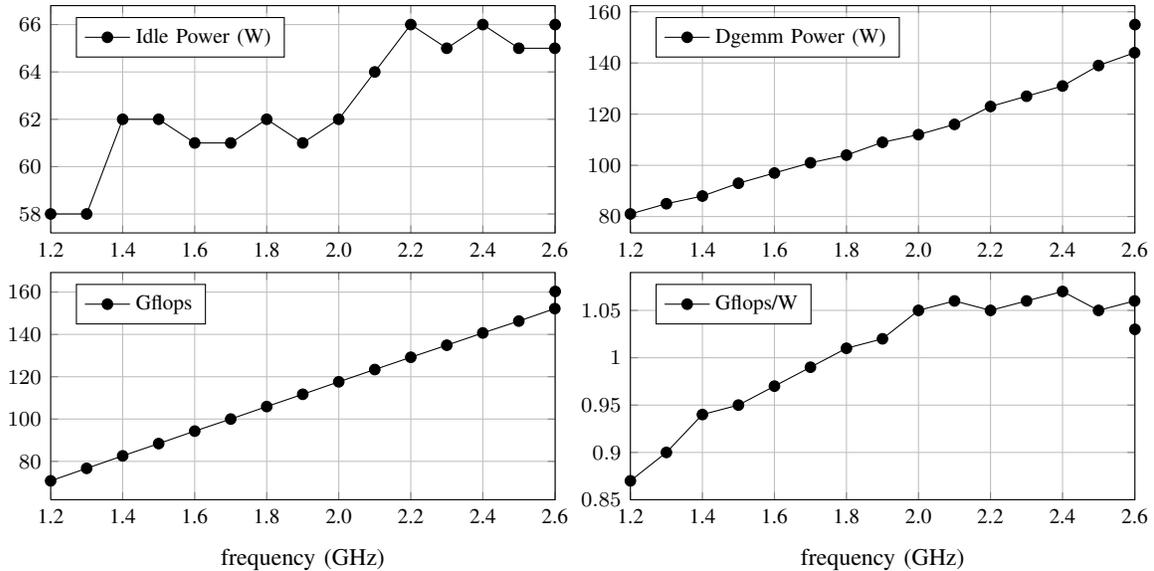
\begin{figure*}[htp!]
    \centering
\begin{tikzpicture}
    \begin{groupplot}[
        group style={
            group name=percolumn,
            group size=2 by 2,
            xlabels at=edge bottom,
            vertical sep=15pt
        },
        footnotesize,
        width=0.45\textwidth,
        height=0.25\textwidth,
        legend style = {at={(0.05,0.95)}, anchor=north west},
        xlabel=frequency (GHz),
        xmin=1.2,xmax=2.601,
        xtick={1.2,1.4,1.6,1.8,2,2.2,2.4,2.6},
        xticklabels={1.2,1.4,1.6,1.8,2.0,2.2,2.4,2.6},
        every axis y label/.style=
            {at={(ticklabel cs:0.5)},rotate=0,anchor=near ticklabel},
        grid = major]
	ylabel=$\text{Gflops/W}$
     \nextgroupplot[]
        \addplot[color=black,mark=*] table[x expr=\thisrow{freq}/1000000,y=idle_power_consumption] {./pstate-perf.tbl};
        \legend{Idle Power (W)};
    \nextgroupplot[]
        \addplot[color=black,mark=*] table[x expr=\thisrow{freq}/1000000,y=dgemm_power_consumption] {./pstate-perf.tbl};
        \legend{Dgemm Power (W)};
    \nextgroupplot[]
        \addplot[color=black,mark=*] table[x expr=\thisrow{freq}/1000000,y=Gflops] {./pstate-perf.tbl};
        \legend{Gflops};
    \nextgroupplot[]
        \addplot[color=black,mark=*] table[x expr=\thisrow{freq}/1000000,y=Gflops/W] {./pstate-perf.tbl};
        \legend{Gflops/W};
    \end{groupplot}
\end{tikzpicture}
\begin{center}
\caption{P-state impact on dgemm performance and node power consumption. The extra value for 2.6 Ghz corresponds to the "turbo boost" frequency, 2.601 Ghz.}
\label{fig:pstate-perf}
\end{center}
\end{figure*}

\subsection{HPL tuning for the Green500 submission}
\begin{figure}[htp!]
    \begin{center}
    \begin{tikzpicture}
    \begin{axis}[
        height=0.3\textwidth,
        width=0.45\textwidth,
        xlabel=$\text{frequency (kHz)}$,
        ylabel=$\text{Gflops/W}$,
        legend style = {at={(0.95,0.05)}, anchor=south east},
        every axis y label/.style=
            {at={(ticklabel cs:0.5)},rotate=90,anchor=near ticklabel},
        grid=major]
        \addplot[color=black,mark=*] table[x=freq,y=Gflops/W] {./HPL_00.tbl};
        \addplot[color=black,mark=*] table[x=freq,y=Gflops/W] {./HPL_99.tbl};
        \addplot[color=black,mark=*] table[x=freq,y=Gflops/W] {./HPL_98.tbl};
        \addplot[color=blue,mark=triangle*, mark size=2] table[x=freq,y=Gflops/W] {./HPL_97.tbl};
        \addplot[color=red,mark=square*] table[x=freq,y=Gflops/W] {./HPL_96.tbl};
        \addplot[color=brown,mark=*] table[x=freq,y=Gflops/W] {./HPL_95.tbl};
        \legend{, ,98--100,97,96,95}
    \end{axis}
\end{tikzpicture}
    \caption{HPL energy efficiency with respect to CPU frequency and DGEMM CPU/GPU split coefficient. The extra value for 2.6 Ghz corresponds to the ``turbo boost'' frequency, i.e. 2.601 Ghz}
    \end{center}
    \label{fig:hpl-perf}
\end{figure}
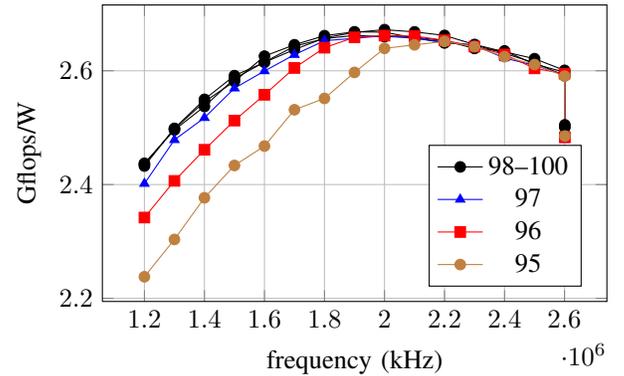

\fig{fig:hpl-perf} shows the performance of the hybrid HPL implementation developed by NVIDIA on 8 nodes of \daint as both the CPU frequency and proportion of work performed on the CPU and GPU are varied.
This figure shows that the best $\mbox{GF/W}$ ratio is achieved at 1.9 Ghz with as much work as possible performed on the GPU. It is interesting to notice that:
\begin{itemize}
\item activating the turbo boost frequency is is very energy inefficient;
\item above 2.0 Ghz, the energy efficiency is roughly the same for all DGEMM CPU/GPU workload ratios;
\item and below 2.0 Ghz, putting more DGEMM work on the CPU will decrease the energy efficiency.
\end{itemize}
Based on those observation, we inferred that the best configuration in order to optimize the $\mbox{GF/W}$ ratio for HPL was to use a DGEMM split ratio of 1 (i.e. perform no DGEMM work on the CPU) and a CPU clock of 2.0 Ghz would be the best starting point for the full green500 runs.

These observations were straightforward to obtain using the PMDB and only 8 nodes, simplifying the initial tuning phase. From these observations, the Green500 setup was as follows
\begin{itemize}
    \item $72\times73$ grid size with 5256 nodes.
    \item N = 3612672.
    \item CPU frequency = 1.9 Ghz.
    \item perform all DGEMM kernels on the GPU.
\end{itemize}

\tbl{tab:green500} shows the power consumption of the Green500 run we submitted to the committee, as measured with the different power measurement capabilities (the level-3 facility readings, which are the most comprehensive, were the actual values submitted). Those measurements are based on the full run in order to do a comparison between job-level measurements and the RUR.

\begin{table}[hp!]
\begin{center}
    \begin{tabular}{|c|r|rr|r|}
        \hline
        CPU freq (Mhz) & RUR &  PMDB (job) & PMDB (cab)  & Facility \\
        \hline
        1.9  & 1526 & 1536 & 1600 & 1635 \\
        \hline
    \end{tabular}
    \vspace{5pt}
    \caption{Power consumption in Watts of the Green500 HPL run using RUR, job-level PMDB, cabinet-level PMDB and Facility power measurements}
\label{tab:green500}
\end{center}
\end{table}

Values from the PMDB and RUR are corrected to include the power consumption from the blowers and the network where needed. \tbl{tab:blowers} shows that the difference between the power consumption of the blowers given by a cabinet-level query of the PMDB compared the power measured with the facility power meters is virtually zero.

From \tbl{tab:green500}, we can see that all of the power consumption measurements are in good agreement.
The small differences between measurements can be explained easily.
The facility power measurement includes all the components of the system, including components like service nodes, IO nodes and disks, which are not included in the RUR the job-level PMDB queries.
However, the service nodes' power consumption is taken into account with the cabinet-level PMDB query.
RUR and job-level PMDB are node-based power measurements and are in very good agreement.
Overall, the differences between any two measurements is within 1\%.

\begin{table}[hp!]
\begin{center}
    \begin{tabular}{|rrr|r|}
        \hline
        PMDB &  PMDB corrected & PMDB Total & Facility \\
        \hline
        4440 & 4673.7 & 79452 & 79448 \\
        \hline
    \end{tabular}
    \vspace{5pt}
    \caption{Blower power consumption given by the cabinet-level PMDB query and the facility power meter in Watts. Note that the PMDB and PMDB results are for an individual blower, and the PMDB and Facility readings are for all 17 blowers.}
    \label{tab:blowers}
\end{center}
\end{table}

\section{Real World Application: Numerical Weather Prediction}
\label{sec:cosmo}
The atmospheric modeling code COSMO (COnsortium for Small-scale MOdeling) was chosen for tests to validate the power measurement capabilities. COSMO is a non-hydrostatic limited area model used for both weather forecasting by national weather services and for climate research.

The production code for COSMO is based on Fortran 90 with a flat MPI parallelization model, optimized for NEC vector architectures.
A Swiss initiative, lead by the Swiss weather service MeteoSwiss, has ported and optimized the entire COSMO code for both multi-core CPU and GPU architectures using CUDA and OpenACC for different parts of the code~\cite{cug2013,gysi2014,cumming2014,lapillonne2014}.
The new GPU implementation is mature, it currently used for production climate simulations on \daint, so it is suitable for comparing both time and energy to solution across different architectures.

In this section we will discuss the configuration for the COSMO simulation that was used for the power measurement tests, followed by a description of the steps taken to ensure that results measured using the PMDB on XC-30 were comparable to those measured with external power meters on other systems. Then a comparison of time and energy to solution across different Cray systems is presented.

\subsubsection{Simulation configuration:}
The configuration used in our tests is based on the 2-km model of the Swiss Alps, known as COSMO-2, used by MeteoSwiss.
A COSMO-2 run uses 9 nodes on each system, 8 nodes for computation, and an additional node for IO.
This model is currently used for daily weather forecasting runs, and it will be used in ensemble mode from 2015.
An ensemble test runs in the order of 20 instances the same simulation (ensemble members), with perturbations to parameters, to better quantify the uncertainty for forecasts.

The ensemble configuration was used for these tests because it allowed us to run enough instances of the 9-node job to fill a cabinet.
It was necessary to fill a cabinet with jobs to compare different systems, because while the XC-30 make it possible to measure power and energy on a node-by-node basis, we were restricted to measuring the power consumption of a whole cabinet with external power meters on the other systems.



\subsubsection{Measurement and systems overview:}
The measurements for time and energy to solution were performed on three systems at CSCS (\rosa, \daint and \todi), and a hybrid XC-30 system named \clogin that we had early access to at Chippewa Falls.
The number of nodes per cabinet on the XE6 and XK7 systems is half of that on the XC-30 systems, so on each system we used as many members in each ensemble run required to fill a full cabinet.

The difference systems used in the tests are summarised below:
\begin{itemize}
\item
    \textbf{\rosa} and \textbf{\todi} are Cray XE6 and XK7 systems respectively, with to 96 compute nodes per rack. Each XE6 node has two by 16-core AMD Interlagos sockets, and the XK7 replaces a socket with a K20X GPU. Each cabinet has a blower integrated in the base. Power measurements were performed at CSCS with an external power meter, which measured total power for one cabinet, including the blower, at a 1Hz frequency.
\item
    \textbf{\daint} and \textbf{\clogin} were Cray XC-30 systems with 192 nodes per cabinet (on \daint 4 of the nodes are service nodes, so one less ensemble member could be run than on \clogin). \daint had two Sandy Bridge sockets with eight cores each per node (this was before \daint's hybrid upgrade), while each node of \clogin had one Sandy Bridge socket with one K20X GPU.
\end{itemize}

In our tests, time and energy to solution were for total time to solution, i.e. measured directly before and after the \textit{aprun} command.
We were interested in the total energy to solution, including the node, network, blower and AC/DC losses in all measurements, which is precisely what is measured by the external power meters on the XE6 and XK7 systems.
The cabinet level readings from the PMDB did not include the blowers on XC-30, so the value for the blowers was approximated according to formula~\eq{eq:energyEstimateWithBlower}, before scaling for AC/DC conversion.

On each system the ensemble runs were performed multiple times to determine the number of runs required to obtain an accurate average results. Both time to solution and energy to solution were highly reproducible with very small variance, so that only two runs were required on each system to have a 99\% confidence level for the mean of the samples to be accurate to 1\%.

\subsection{Results}

Before using results from the PMDB to compare energy to solution on the XC-30 against the other systems, we first validated that the results from the PMDB could be used to derive measurements equivalent to those from an external power meter.
The \clogin system had two external power meters: the first measured power for one cabinet; and the second measured power ``at the wall'' for the whole system, i.e. for all three cabinets and two blowers.

\begin{table}[hp!]
\centering
\begin{tabular}{|l|rrr|}
            \cline{2-4}
\multicolumn{1}{c|}{} & PMDB (kWh)& external (kWh)& efficiency (PMDB/external)\\
            \hline
run 1                 & 53.63         & 56.45         & 95.00\%    \\
run 2                 & 53.47         & 56.27         & 95.02\%    \\
            \hline
\end{tabular}
\vspace{5pt}
\caption{ETS measured using PMDB and external meter for a three-cabinet hybrid XC-30 system \clogin. The energy is for all three cabinets with the blowers, with the blower power assumed to be \watt{4440} for the PMDB results (see \tbl{tab:blowers}).}
\label{tbl:cosmoValidate}
\end{table}

An ensemble run large enough to fill all three cabinets was run two times, and results from both power meters were compared to those from the PMDB.
The PMDB agreed very well with the power meter for the entire system with the assumed 95\% efficiency in the conversion from AC to DC, as shown in the results in \tbl{tbl:cosmoValidate}.
The power meter on the single cabinet disagreed significantly, by around 5\% with the PMDB, however further investigation showed that this meter required recalibration, and that the PMDB results were correct.

\begin{table}[htp!]
\centering
\begin{tabular}{|l|rrcc|}
\hline
System                  & XE6       & XK7       & XC-30      & hybrid XC-30   \\
Name                    & \rosa     & \todi     & \daint    & \clogin       \\
Configuration           & CPU       & \textbf{GPU} & CPU    & \textbf{GPU}  \\
\hline
Ensemble Size           & 10        & 10        & 20        & 21            \\
       \rowcolor{lightred}
Wall Time (s)           & 3683      & 2579      & 2083      & 1539          \\
Mean Power (kW)         & 40.22     & 62.07     & 28.27     & 41.56         \\
Peak Power (kW)         & 43.00     & 64.96     & 31.55     & 43.97         \\
Energy to solution (kWh)& 41.14     & 44.47     & 16.36     & 17.77         \\
       \rowcolor{lightblue}
Energy per member (kWh) & 4.11      & 2.22      & 1.636     & 0.85          \\
\hline
       \rowcolor{lightred}
Speedup time to solution& 1.0       & 1.42      & 1.76      & 2.39          \\
       \rowcolor{lightblue}
Speedup energy to solution & 1.0    & 1.85      & 2.51      & 4.84          \\
\hline
\end{tabular}
\vspace{5pt}
\caption{Results of time to solution and energy to solution for COSMO-2 ensemble runs on different Cray systems.}
\label{tbl:cosmoResults}
\end{table}

\tbl{tbl:cosmoResults} shows the result for all of the Cray systems. There are two clear trends in terms of architecture. The first is that moving from one generation of system to another, from XE6 to XC-30 and from XK7 to hybrid-XC-30, improves both TTS and ETS in all cases. This makes sense, due to improvements including CPUs (Interlagos to Sandy Bridge) and network (Gemini to Aires). The second trend is that TTS and ETS are also better on GPU-based systems. It is interesting to note that the ETS improvements are greater than TTS improvements in all cases: indeed energy to solution per ensemble member is a factor of 4.8, while TTS improves by a factor of 2.4, going from XE6 to hybrid XC-30.

The original aim of this investigation was to use external power meters for all systems, to provide comprehensive energy results.
We had the opportunity to validate the PMDB counters with early access to a hybrid XC-30 system, and the results from those experiments, presented above gave us confidence that the built in power monitoring features can be used to accurately measure the true energy to solution for applications.
Measurements with the PMDB were certainly much easier than using external meters, and the PMDB has become the go to tool for follow on work with COSMO~\cite{cumming2014}.

\section{Real World Application: dca++}
\label{dca}
DCA++ is a simulation code designed to solve Dynamic Cluster Approximation (DCA) models of high-temperature superconductors, such as the two-dimensional Hubbard model and extensions that allow studies of disorder effects and nanoscale inhomogeneities.
It is based on a continuous time quantum Monte-Carlo solver with delayed updates which allows to use an efficient algorithm based on BLAS level 3 operations (DGEMM) that are mostly computed on the GPU. DCA++ is therefore capable of reaching above 50\% of the peak performance\cite{Staar:2013:TQL:2503210.2503282}.

We measured the power consumption for several test cases using the PMDB features and an external power meter on the hybrid XC-30 clogin at Chipewa falls.
Each run corresponds to a different temperature (in Kelvin) expressed in terms of $t$, the Hopping parameter, assumed to be the 1 electron-volt in the calculations.
This temperature is represented by the beta coefficient in the input file: for beta between 5 and 35,  
\begin{equation*}
T = \frac{11604}{\mbox{beta}}.
\end{equation*}
For beta equal to 40, we reach the cuprate temperature and must adjust the formula to obtain a temperature in Kelvin in the range of the high-Tc cuprate family:
\begin{equation*}
T = \frac{11604}{\mbox{beta}}*0.25 = 72.5.
\end{equation*}

\tbl{tbl:dca-results} shows the average power consumption measured by cabinet-level PMDB queries and an external power meter with a $0.1 \mbox{Hz}$ smapling frequency.

\begin{table}[htp!]
\centering
\begin{tabular}{|l|rrrrrrrr|}
\hline
Beta                      & 5 & 10 & 15  & 20 & 25 & 30 & 35 & 40 \\
TTS (s)      & 3787 & 2725 & 922 & 605 & 329 & 182 & 75 & 23 \\
\hline
cab PMDB (kW)       & 58.5 & 57.2 & 55.9 & 53.6 & 53.2 & 49.3 & 46.8 & 43.0 \\
External PM (kW) & 58.6 & 57.1 & 53.9 & 52.9 & 52.2 & 47.0 & 42.7 & 30.5 \\
\hline
\end{tabular}
\vspace{5pt}
\caption{Power consumption of DCA+ test cases measured by the cabinet-level PMDB and an external power meter at Chipewa Falls}
\label{tbl:dca-results}
\end{table}

Results are in very good agreement for the test cases with large Time To Solution. For small TTS, the external power meter resolution (10 seconds) is not high enough to give a reliable measurement.

\section{Conclusion}
\label{sec:conclusion}
In this paper, we have shown direct comparisons between the newly-introduced power measurement data base (PMDB) and both the Level 3 capable facility power meters available at CSCS and standard power meters directly attached to an hybrid XC-30 prototype. Results are in very good agreement and show that the PMDB can reliably be used to measure power consumptions of applications running on the XC-30 systems. The PMDB tool provides a power measurements granularity (node, accelerator, blades racks) that standard power meters cannot provide easily and therefore this tool is a perfect complement to the facility power measurement tools. Since most of the PMDB informations are available in user space, optimization and tuning processes using multi-dimensional parameters search spaces, such as CPU/GPU frequency and overlapping of computations, for energy efficiency become easier and more straightforward, as shown for the Green500 tuning.

\section*{Acknowledgment}

The authors would like to thank all the people who helped with taking energy measurements.
In particular, people from Cray in Chippewa Falls  Steve Martin and Ron Rongstad introduced us to the PMDB, and helped with manual power measurement.
Nina Suvanphim provided assistance on site at CSCS. Finally Tiziano Belotti, Rolando Summermatter and Luca Bacchetta from the Facility Management Group at CSCS performed all of the external power monitoring. Finally, Massimiliano Fatica from nVidia who provided the hybrid HPL code which was used for the Green500 submission. 


\bibliographystyle{abbrv}
\bibliography{paper}

\begin{thebibliography}{10}

\bibitem{eehpcwg:website}
{Energy Efficient High Performance Computing Working Group} homepage, {URL:
  \texttt{eehpcwg.lbl.gova}}.

\bibitem{netlib:website}
{Netlib HPL benchmark} homepage, {URL: \texttt{www.netlib.org/benchmark/hpl/}}.

\bibitem{green500:website}
{The Green500 List} homepage, {URL: \texttt{www.green500.org}}.

\bibitem{craypmdb}
Monitoring and managing power consumption on the cray xc30 system.
\newblock Technical report, 2014.

\bibitem{cumming2014}
B.~Cumming, G.~Fourestey, O.~Fuhrer, M.~Fatica, and T.~C. Schulthess.
\newblock Application centric energy-efficiency study of distributed multi-core
  and hybrid {CPU-GPU} systems.
\newblock \emph{submitted} SC '14.

\bibitem{cug2013}
B.~Cumming, C.~Osuna, T.~Gysi, X.~Lapillonne, O.~Fuhrer, and T.~C. Schulthess.
\newblock A review of the challenges and results of refactoring the community
  climate code {COSMO} for hybrid cray {HPC} systems.
\newblock {\em Cray User Group}, 2013.

\bibitem{LINPACKuserguide}
J.~Dongarra, J.~Bunch, C.~Moler, and G.~Stewart.
\newblock {\em LINPACK Users Guide}.
\newblock {SIAM}, Philadelphia, {PA}, 1979.

\bibitem{gysi2014}
T.~Gysi, O.~Fuhrer, C.~Osuna, M.~Bianco, and T.~Schulthess.
\newblock {STELLA}: A domain-specific language and tool for structured grid
  methods.
\newblock \emph{submitted} SC '14.

\bibitem{lapillonne2014}
X.~Lapillonne and O.~Fuhrer.
\newblock Using compiler directives to port large scientific applications to
  {GPUs}: An example from atmospheric science.
\newblock {\em Parallel Processing Letters}, 24(01):1450003, 2014.

\bibitem{scogland2014}
T.~R. Scogland, C.~P. Steffen, T.~Wilde, F.~Parent, S.~Coghlan, N.~Bates, W.-c.
  Feng, and E.~Strohmaier.
\newblock A power-measurement methodology for large-scale, high-performance
  computing.
\newblock In {\em Proceedings of the 5th ACM/SPEC International Conference on
  Performance Engineering}, ICPE '14, pages 149--159, New York, NY, USA, 2014.
  ACM.

\bibitem{Staar:2013:TQL:2503210.2503282}
P.~Staar, T.~A. Maier, M.~S. Summers, G.~Fourestey, R.~Solca, and T.~C.
  Schulthess.
\newblock Taking a quantum leap in time to solution for simulations of high-tc
  superconductors.
\newblock In {\em Proceedings of the International Conference on High
  Performance Computing, Networking, Storage and Analysis}, SC '13, pages
  1:1--1:11, New York, NY, USA, 2013. ACM.

\end{thebibliography}

\end{document}